\documentclass[aps,pre,twocolumn]{revtex4-2}

\usepackage{graphicx}
\usepackage{amsmath}
\usepackage{amssymb}
\usepackage{braket}
\usepackage{physics}
\usepackage{bbold}
\usepackage{hyperref}
\usepackage{subfigure}
\usepackage[usenames,dvipsnames]{color}
\usepackage{graphicx}
\usepackage{subfigure}
\usepackage[mathscr]{euscript}

\begin{document}

\title{Periodically driven many-body quantum battery}
\author{Saikat Mondal}
\email{msaikat@iitk.ac.in} 
\author{Sourav Bhattacharjee}
\email{bsourav@iitk.ac.in}
\affiliation{Department of Physics, Indian Institute of Technology Kanpur, Kanpur 208016, India}
%\date{\today}

\begin{abstract}
We explore the charging of a quantum battery based on spin systems through periodic modulation of a transverse-field like Ising Hamiltonian. In the integrable limit, we find that resonance tunneling can lead to a higher transfer of energy to the battery and better stability of the stored energy at specific drive frequencies. When the integrability is broken in the presence of an additional longitudinal field, we find that the effective Floquet Hamiltonian contains terms which may lead to a global charging of the battery. However, we do not find any quantum advantage in the charging power, thus demonstrating that global charging is only a necessary and not sufficient condition for achieving quantum advantage.  
\end{abstract}
\maketitle

\section{Introduction}\label{sec_intro}
In recent times, there has been a growing interest towards the understanding of thermodynamic aspects of small-scale quantum systems \cite{breuer, gemmer10, kosloff13,sai16, deffner19, kosloff19}. This has necessitated a closer look at the dynamics of energy conversion at such small scales, with a particular focus on how useful work can be extracted from thermal sources utilizing what are known as quantum thermal machines~\cite{mukherjee_review,bhattacharjee_review}. As an offshoot of this development, efforts have also been made to explore the possibility of efficiently storing the extracted energy in so-called `quantum batteries'~\cite{alicki13, binder_review, bhattacharjee_review}, with the expectation that quantum effects may lead to a superior performance of such batteries when compared to their classical counterparts \cite{campaioli17, ferraro18, rossini19, farina19, farre20, rossini20, rosa20, bai20, caravelli20, caravelli21, salvatore21}. 
%In particular, the   batteries. Quantum thermal machines (QTMs) are microscopic devices that are quantum analogues of classical thermal machines (e.g., Carnot engine, Otto engine etc.),
%These machines perform their operations with the transfer of heat energy. 
 %though laws of quantum mechanics play  an important role in the operation of QTMs.
% The laws of thermodynamics in quantum regime~\cite{kosloff13} lead to some interesting phenomena in these QTMs. 

The performance of a quantum battery is usually judged on three aspects -- the maximum amount of extractable energy, charging/discharging power and the stability of the stored energy. In the simplest of settings where environmental dissipation can be ignored, the charging or the transfer of energy to the battery is carried out using a unitary quench protocol. In other words, given that the quantum system designated as the battery is initially in the ground-state of the `battery Hamiltonian', the charging process involves turning on additional interactions or external fields for a finite time during which the state of the battery evolves unitarily, driven by a net `charging Hamiltonian'. At the end of the charging process, the additional interactions/fields are switched off; the energy difference between the final and initial state with respect to the battery Hamiltonian is considered as the energy stored in the battery. However, in the reverse process, it is often not possible to extract all the energy stored in the battery through unitary protocols, and the maximum amount of the useful energy that can be extracted is termed as ergotropy \cite{binder_review}. The rate at which energy is stored (extracted) in the battery is defined as its charging (discharging) power which should ideally be maximized. Finally, a high variance in the charged state of the battery with respect to the battery Hamiltonian implies an undesirable instability in the energised state of the battery \cite{friis18}.

For a battery composed of $N$ identical quantum systems or `cells', parallel charging operations -- where each of the cell is separately charged -- can only lead to a linear scaling of the charging power with $N$. On the contrary, collective charging protocols in which the charging Hamiltonian couples multiple cells simultaneously can lead to a super-linear scaling of the charging power with the system size; a \textit{quantum advantage} \cite{campaioli17} is thus said to emerge in this scenario. Recently, it has been shown that global charging protocols, in which all the cells are simultaneously charged, leads to the maximal possible quantum advantage where the charging power scales quadratically with system size \cite{rosa21}. However, it has also been hinted that such protocols are only a necessary condition at best and not a sufficient one. \cite{binder15,farre20,rossini20,rosa21}.  

In this work, we depart from the traditional way of charging a quantum battery through a quench protocol and consider the case of a periodic driving protocol. To elaborate, we choose the charging Hamiltonian to be a periodic function of time and the energetics of the battery is observed only at stroboscopic instants which are separated by an interval equal to the time period of the charging Hamiltonian. It is well known that under such a periodic driving protocol, the dynamics of the system at stroboscopic instants is dictated by a time-independent Floquet Hamiltonian \cite{bukov15,eckardt17}, which can have drastically different properties from that of the charging Hamiltonian. As we shall demonstrate, this leads to the emergence of interesting features in the performance of the quantum battery. 

Although quantum batteries can be modeled in a variety of ways, we consider a simple model based on a lattice system of $N$ non-interacting half-integer spins. Firstly, we focus on the case where the battery is charged using a periodically modulated transverse-field Ising Hamiltonian (TFIH) \cite{sachdev11,suzuki13,dutta15}with nearest-neighbour interactions. We find that the energy transferred to the battery can be enhanced depending on the frequency of the charging Hamiltonian. However, the translationally invariant integrable TFIH is incapable of exhibiting a quantum advantage in the charging power. This is not surprising as the TFIH is equivalent to a lattice model of non-interacting free-fermions with local decoupled structure in the momentum space \cite{farre20}. The local structure effectively allows the TFIH to be represented as a collection of N decoupled pseudo-spins, thus implying a parallel charging scenario. We note here that integrability by itself is not necessarily a hurdle to achieving quantum advantage, rather it is the existence of local decoupled structure which allows at most a linear scaling of the charging power. In this regard, it has been shown that quantum advantage can emerge for certain free variations of the Sachdev-Ye-Kitaev model, which are integrable~\cite{rosa22}.
	
To overcome this limitation in the charging power arising from the effective local structure of the TFIH, we next consider the case where an additional periodically modulated longitudinal field is used to charge the battery, along with the TFIH. This breaks the integrability of the model and no longer permits a description of the system in terms of decoupled pseudo-spins. In this case, we find that even though the effective Floquet Hamiltonian approaches a global charging like scenario at low drive frequencies, no quantum advantage is seen to emerge in the charging power.

The rest of the paper is organized as follows: in Sec.~\ref{sec_periodic_driving}, we introduce the periodically driven quantum battery and show how the relevant quantities are to be calculated. In Sec.~\ref{sec_Ising}, we investigate a quantum battery driven with the integrable TFIH and show that the asymptotic energy stored in the battery strongly depends on the drive frequency. In Sec.~\ref{sec_longitudinal_field}, we proceed to explore the charging power of the battery in presence of an additional integrability-breaking longitudinal field, having same periodicity as that of the TFIH. Concluding remarks are provided in Sec.~\ref{sec_summary}.

\section{Periodically driven quantum battery}\label{sec_periodic_driving}

We consider a quantum battery composed of $N$ non interacting half-integer spins in presence of a magnetic field, represented by the Hamiltonian, 
\begin{equation}\label{eqn_H_B}
H_{B} = h_{z} \sum_{j=1}^{N} \sigma_{j}^{z} ,
\end{equation}
where $\sigma_{j}^{z}$'s are Pauli matrices. During the charging process, the state of the battery evolves under the charging Hamiltonian, $H_{c}(t) = H_{B} + V(t)$, for a duration of time $\tau$, such that $V(0)=V(\tau)=0$. At $t=0$, the battery is assumed to be in the ground state $\rho(0)=\ket{\psi_0}\bra{\psi_0}$ of the Hamiltonian $H_{B}$. 
%The density matrix of the battery at $t=0$ is $\rho(t=0) = \rho_{0} = \ket{\psi (0)} \bra{\psi (0)}$.
%
%The time evolution of the density matrix is governed by Liouville-Von Neumann equation~\cite{campaioli18}
%\begin{equation}\label{eqn_d_rho_dt}
%\frac{d \rho (t)}{dt} = - i [ H_{c}(t), \rho (t) ] ,
%\end{equation}
%where we have used natural unit (i.e. $\hbar=1$). The solution of eqn.(\ref{eqn_d_rho_dt}) is of the following form:
At the end of the charging process, the time-evolved state of the battery is given by,
\begin{equation}\label{eqn_rho_t}
\rho (\tau) = U (\tau) \rho_{0} U^{\dagger} (\tau) ,
\end{equation}
where $U (\tau) = \mathcal{T} \exp (-i \int_{0}^{\tau} H_{c}(t^{\prime}) dt^{\prime})$ with $\mathcal{T}$ being the time ordering operator. Note that we have set $\hbar=1$. 
In what follows, we shall consider a periodically driven battery such that,
\begin{equation}\label{eqn_H_c_periodic}
H_{c} (t) = H_{c} (t + T) ,
\end{equation}
where $ T = 2 \pi / \omega$ is the time period of the charging Hamiltonian and $\omega$ is the frequency of the drive. Naturally, this requires $V(t+T)=V(t)$. Further, we shall restrict ourselves to the stroboscopic dynamics of the charging process; the charging time $\tau$  therefore shall always satisfy the condition $\tau=nT$, where $n\in\mathbb{Z}^+$. At these instants, the time-evolution operator assumes the form, $U(nT)=(U^F)^n$, where the Floquet operator $U^F$ is given by \cite{bukov15, eckardt17},
\begin{equation}\label{eqn_Floquet}
U^F = \mathcal{T} \exp (-i \int_{0}^{T} H_{c}(t^{\prime}) dt^{\prime}) = \exp (-i H_{c}^{F} T).
\end{equation}
In other words, when observed only at the stroboscopic instants, the battery evolves under the action of the time-independent Floquet Hamiltonian $H_{c}^{F}$.

We now define a set of quantities which will be relevant for rest of the paper. The energy stored in the battery after $n$ stroboscopic instants is given by,
\begin{align}\label{eqn_E_t}
	E(n)& = \Tr \left[\rho (nT) H_{B}\right] - \Tr \left[\rho_{0} H_{B}\right] \nonumber \\
	& = \Tr \left[(U^F)^n \rho_{0} (U^F)^{n\dagger} H_{B}\right] - \Tr \left[\rho_{0} H_{B}\right],
\end{align}
and the charging power is obtained as,
\begin{equation}\label{eqn_nst_P}
	P(n) = \frac{E(n)}{nT}.
\end{equation}
%At stroboscopic time $t = n T$ $ \forall n \in \mathbb{Z_{+}}$, the state of the system is determined by
%\begin{equation}\label{eqn_psi_t}
%\ket{\psi (n T)} = U^{n} (T) \ket{\psi (0)} .
%\end{equation}
%The average energy stored over the span of time  $\tau = q T$ with $q \in \mathbb{Z_{+}}$ and $q>>1$ turns out to be
%\begin{equation}\label{eqn_E_average}
%\overline {E(\tau)} = \frac{1}{q} \int_{0}^{q} E(t=n T) dn ,
%\end{equation}
%where $E(t=n T)$ can be evaluated as follows:
%\begin{equation}\label{eqn_E_nT}
%E (t=n T) = \expval{H_{B}}{\psi (n T)}-\expval{H_{B}}{\psi (0)}.
%\end{equation}
%The average charging power is  calculated as,
%\begin{equation}\label{eqn_average_P}
%P = \frac{\overline{E (\tau)}}{\tau}.
%\end{equation}
Similarly, the variance of the battery Hamiltonian at time $t=nT$ is determined by
\begin{equation}\label{eqn_var_H_B}
\Delta H_{B}^{2}(n) = \Tr\left[\rho(nT)H_{B}^{2}\right]-\Tr\left[\rho(nT)H_{B}\right]^2.
\end{equation}

%In the upcoming sections, we shall explore periodic driving of the quantum battery with two different charging Hamiltonians and the variation of charging power with frequency of periodic driving. We shall also compare the scaling of charging power with the number of TLSs $N$ in these two cases and look for super-extensive scaling of charging power. 
%In case of super-extensive scaling of power, we should get, $P \sim N^{x}$ with $x>1$. For extensive scaling, $x=1$ (i.e. $P \sim N$). 
%Fast charging of quantum battery is possible if we can find super-extensive scaling of power {\cite{farre20}}.

\section{Periodic driving with integrable transverse-field Ising Hamiltonian}\label{sec_Ising}
\begin{figure}
	\centering
		\includegraphics[width=0.45\textwidth]{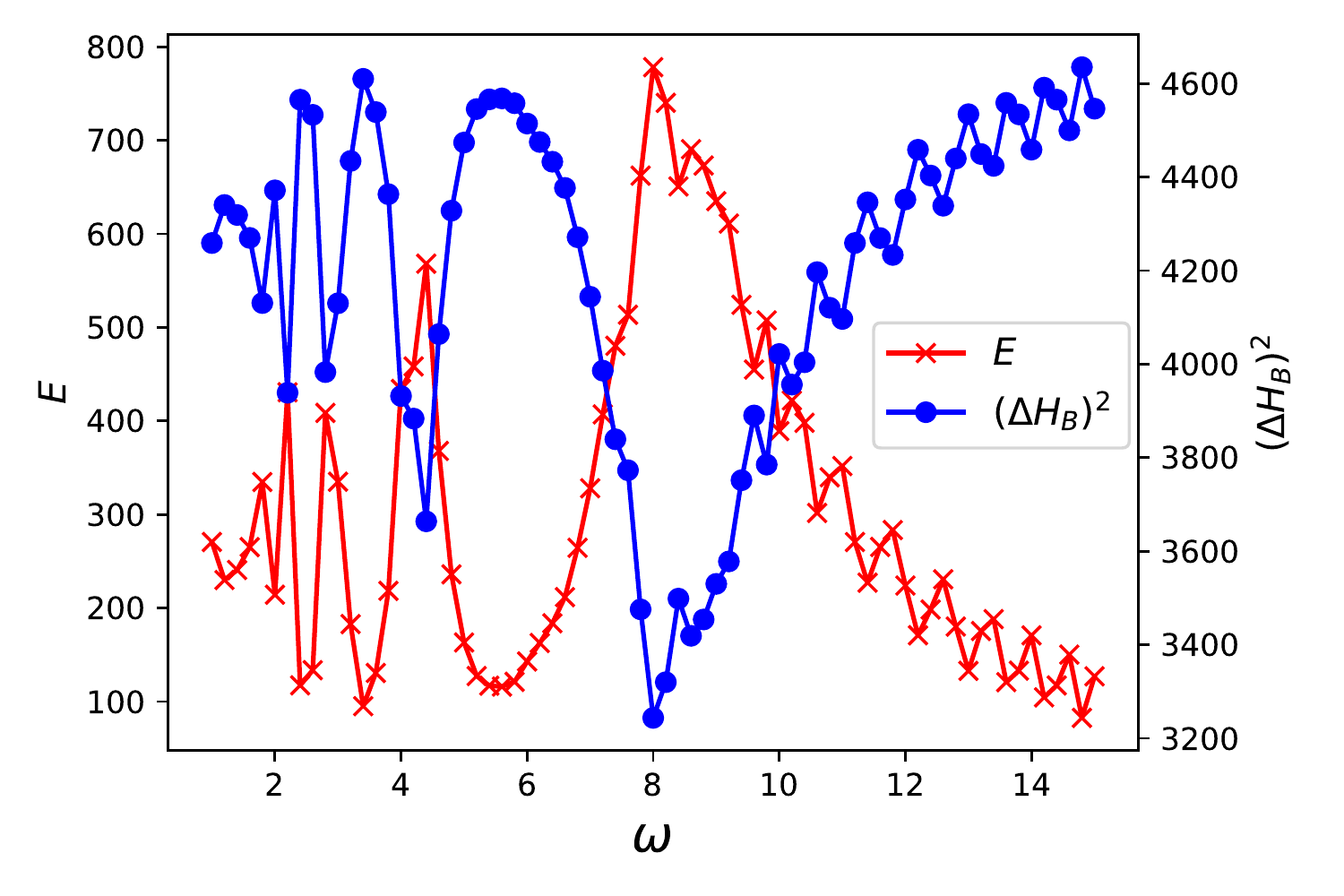}	
	\centering
	\caption{ Energy ($E$) stored in the battery and the variance of the battery Hamiltonian $(\Delta H_{B})^{2}$ as a function of the drive frequency after $n=100$, when the battery is charged using the integrable TFIH (see Eq.~\eqref{eqn_Hc_Ising}). The other parameters chosen for numerics are $J_{0}=1$, $h_{z}=2$ and $N=200$.}\label{fig_freq_int}.
\end{figure}

In this section, we consider a periodic modulation to the battery Hamiltonian of the form,
\begin{equation}\label{eqn_Hc_Ising}
V(t) = J(t) \sum_{j=1}^{N-1} \sigma_{j}^{x} \sigma_{j+1}^{x},
\end{equation} 
so that the charging Hamiltonian $H_c(t) = H_B + V(t)$ becomes identical to that of the TFIH. For the drive, we choose a square-pulse modulation,
\begin{equation}\label{eqn_J_t}
J (t) = \begin{cases}
J_{0} , & \text{for $ nT < t < \left(n+\frac{1}{2}\right)T $}  \\
-J_{0}  , & \text{for $ \left(n+\frac{1}{2}\right)T < t < \left(n+1\right)T $}.
\end{cases}
\end{equation}
%In the context of quantum Ising model, $h_{z}$ can be thought of a static transverse field. 
The TFIH is integrable and becomes analytically tractable after a Jordan-Wigner mapping \cite{lieb61,dutta15} to an equivalent model of non-interacting spinless-fermions and a further assumption of translational invariance. A detailed analysis of the same is given in Appendix.~\ref{app_ising_integrable}. In the quasi-momentum space, the charging Hamiltonian acquires a simple decoupled structure, which is also manifested in the form of the Floquet operator,
% 
%\begin{equation}\label{eq_hc_momentum}
%	H_c(t) = \bigoplus_{k\geq0} H_{c,k}(t).
%\end{equation}
%The Floquet operator is found to be of the form,
\begin{widetext}
\begin{equation}\label{eqn_U_k_T}
	U^F = \bigotimes_{k\geq0}U^F_{k}=\bigotimes_{k\geq0}e^ {- \frac{iT}{2} \left( 2 J_{0} \sin(k) \eta_{y} + \left( 2 h_{z} - 2 J_{0} \cos(k) \right) \eta_{z} \right) }
	e^ {- \frac{iT}{2} \left( - 2 J_{0} \sin(k) \eta_{y} + \left( 2 h_{z} + 2 J_{0} \cos(k) \right) \eta_{z} \right)}, 
\end{equation}
\end{widetext}
where $\eta_{y}$ and $\eta_{z}$ are Pauli matrices representing pseudo-spins. 

The energy stored in the battery as well as its variance are found to saturate to steady values in the limit $n\to\infty$ and $N\to\infty$, when observed at stroboscopic intervals. In Fig.~\ref{fig_freq_int}, the values of these quantities  after sufficiently long time ($n=100$) are plotted as a function of the drive frequency ($\omega$) for $N=200$. It is clear that certain drive frequencies favour a much higher storage of energy in the battery. In addition, the minima of the variance are found to coincide with the energy maxima (see Appendix~\ref{app__int} for explanation), implying that the energy stored at these frequencies also has higher stability. The extrema in the energy and variance arise due to resonance tunneling  -- the eigenspectrum of the Floquet operator becomes degenerate \cite{russomanno12}. This is easily seen from Eq.~\eqref{eqn_U_k_T}, where we note that  $U_{k = 0}^{F} = U_{k = \pi}^{F} = \mathbb{1} $, for $\omega = 2 h_{z}/p$, with $p \in \mathbb{Z}^+$. Similarly,  $U_{k = 0}^{F} = U_{k = \pi}^{F} = - \mathbb{1} $ for $\omega = 4 h_{z}/(2 p +1)$. Thus, it is evident that the drive frequency in a periodically driven quantum battery provides an additional control, tuning which one can have a precise control over the amount of energy stored in the battery and its stability during the charging process.

It is well known that a battery of spins driven with a static TFIH exhibits no quantum advantage in charging power \cite{farre20}. The same can be expected to hold true for the periodically driven case. When observed at stroboscopic instants, the evolution of the battery is equivalent to that of the evolution driven by the static Hamiltonian $H_c^F$. Thus, following Ref.~[\onlinecite{farre20}], the instantaneous charging power at stroboscopic instants is upper bounded as,
\begin{equation}\label{eqn_bound}
	P_{ins}(n)\leq\sqrt{\Delta H_B^2(n)\Delta H_c^2(n)},
\end{equation}
where $\Delta H_B^2(t)$ and $\Delta H_c^2(t)$ are the instantaneous variance of the evolved state with respect to the battery and charging Hamiltonians, respectively. A genuine quantum advantage in the instantaneous charging power is said to emerge (also reflected in the average charging power defined in Eq.~\eqref{eqn_nst_P}) if $\Delta H_B^2(t)$ scales super-extensively with the system size, which in turn requires long-ranged entanglement to be generated between the spins. However, as already mentioned above, that the spin chain battery driven with the TFIH is equivalent to a chain of decoupled pseudo-spins, with each of the pseudo-spins being charged independently. Consequently, each of the variances in Eq.~\eqref{eqn_bound} can also scale at most linearly with the system size (see Appendix.~\ref{app__int} for more details).  Hence the integrability of the TFIH rules out any possible quantum advantage.  

\section{Periodic driving with transverse-field Ising Hamiltonian and a longitudinal field}\label{sec_longitudinal_field}
\begin{figure}
	\centering
	\includegraphics[width=0.45\textwidth]{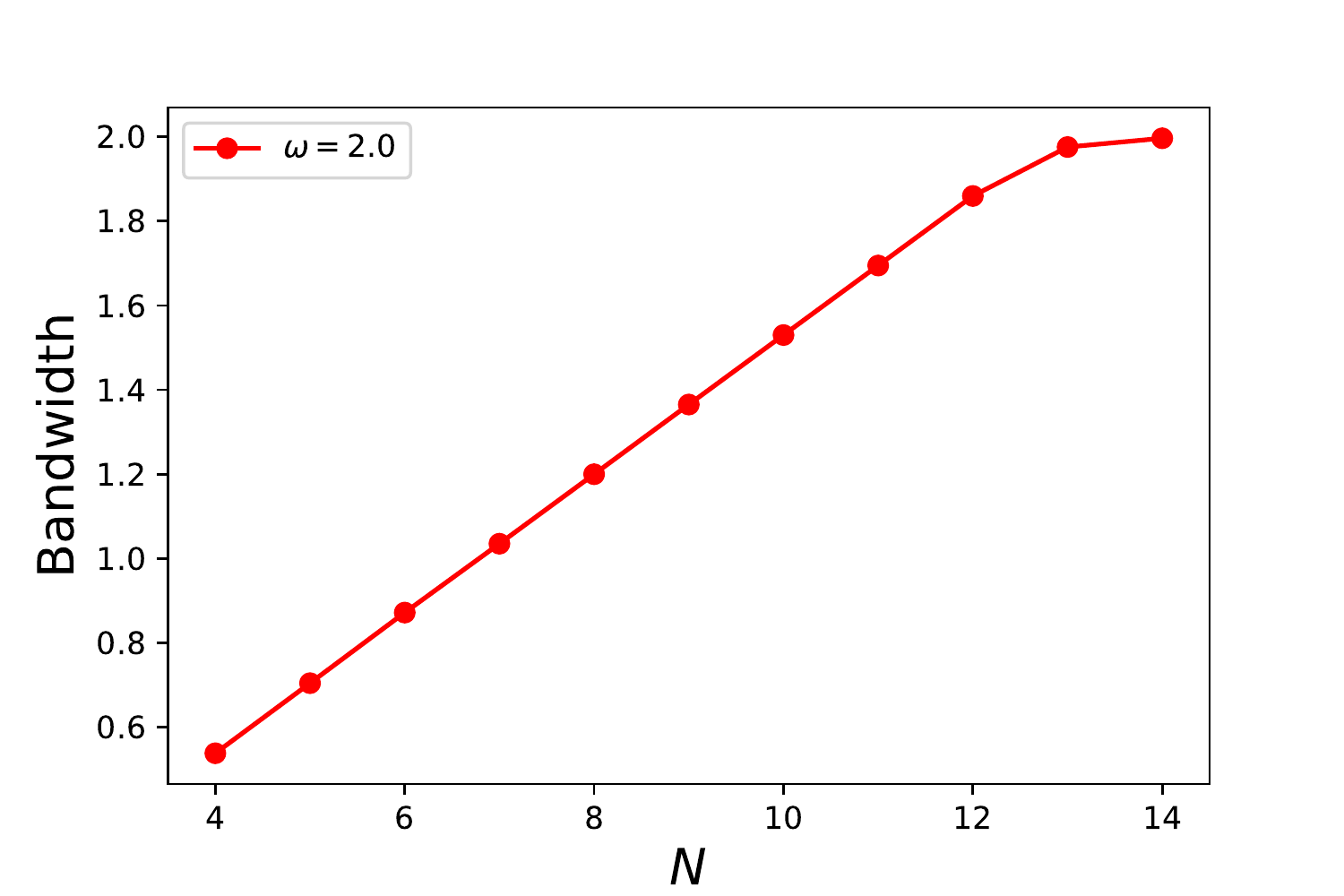}	
	\centering
	\caption{Bandwidth (difference between the largest eigenvalue and the smallest eigenvalue) of Floquet Hamiltonian ($H_{c}^{F}$) as a function of the number ($N$) of spin-$1/2$ systems, when the battery is charged using a modulation of the form given in Eq.~\eqref{eqn_Hc_hx} with the drive-frequency $\omega=2.0$.} \label{fig_bandwidth}.
\end{figure}

In this section, we investigate the charging of the same quantum battery  but with a non-integrable charging Hamiltonian. Specifically, the charging Hamiltonian is chosen to be that of the TFIH with an additional longitudinal field;  the latter having the same time-periodicity as that of the TFIH. In other words, we consider a modulation of the form,
\begin{equation}\label{eqn_Hc_hx}
V(t) =  J(t) \sum_{j=1}^{N} \sigma_{j}^{x} \sigma_{j+1}^{x} + h_{x}(t) \sum_{j=1}^{N} \sigma_{j}^{x} ,
\end{equation} 
where $J(t)=J(t+T)$ and $h_{x}(t)=h_{x}(t+T)$. Once again, we consider a square-pulse charging protocol with $J(t)$ being the same as given in Eq.~\eqref{eqn_J_t} and  $h(t)$ given by,
\begin{equation}\label{eqn_hx_t}
h_{x} (t) = \begin{cases}
 h_{0} , & \text{for $ nT < t < \left(n+\frac{1}{2}\right)T $}\\
- h_{0} , & \text{for $ \left(n+\frac{1}{2}\right)T < t < \left(n+1\right)T $}.
\end{cases}
\end{equation}
Clearly the integrable limit discussed in Sec. \ref{sec_Ising} is retrieved when $h_{0}=0$.

\begin{figure}
	\centering
	\includegraphics[width=0.45\textwidth]{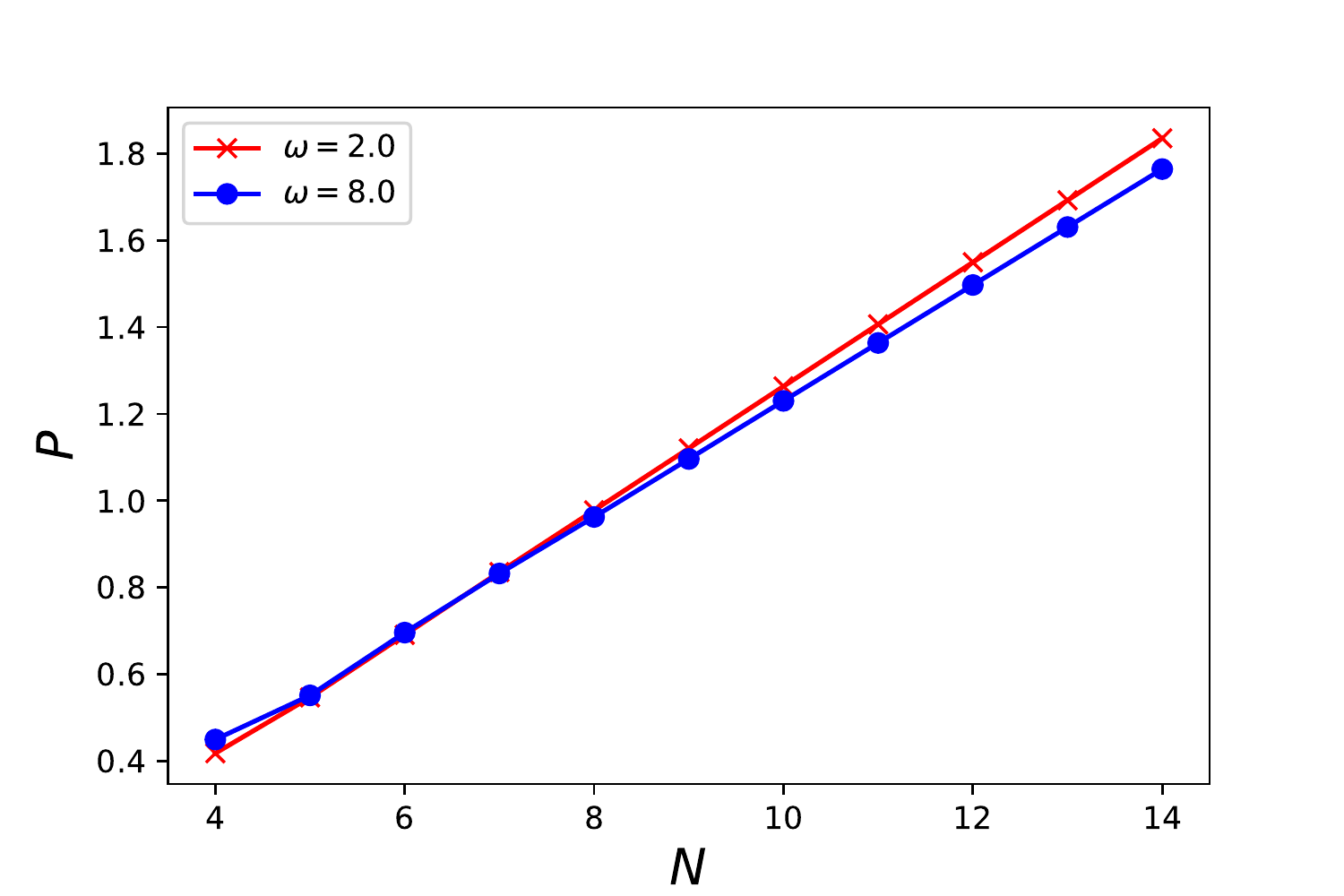}	
	\centering
	\caption{ Charging power ($P$) as a function of the number ($N$) of spin-$1/2$ systems after $n=n^{*}$ for two different drive frequencies ($\omega=2$ and $\omega=8$), when charged using a modulation of the form given in Eq.~\eqref{eqn_Hc_hx}. Here, $n^{*}$ is the number of time periods for which $P$ becomes maximum. The other parameters chosen for numerics are $J_{0}=0.5$, $h_{z}=2$ and $h_{0}=0.3$.}\label{fig_n_p}
	%For $\omega=2$ and $\omega=8$, the values of $n^{*}$ are found to be $2$ and $3$,respectively..
\end{figure}

Firstly, we note that the charging Hamiltonian can no longer be mapped to the model of decoupled pseudo-spins. Next, let us take a look at the explicit form of the corresponding Floquet Hamiltonian $H_{c}^{F}$,
\begin{multline}\label{eq_Floquet}
H_{c}^{F} \approx h_{z} \sum_{j} \sigma_{j}^{z} + \frac{T}{2} \Big[ h_{z} J_{0} \sum_{j} \sigma_{j}^{x} \sigma_{j+1}^{y} + h_{z} J_{0} \sum_{j} \sigma_{j}^{y} \sigma_{j+1}^{x} \\ 
+ h_{z} h_{0} \sum_{j} \sigma_{j}^{y} \Big]
- \frac{T^{2}}{3} \Big[ h_{0} J_{0} h_{z} \sum_{j} \sigma_{j}^{z} \sigma_{j+1}^{x} + h_{0} J_{0} h_{z} \sum_{j} \sigma_{j}^{x} \sigma_{j+1}^{z} \\ 
+ h_{z} \left( J_{0}^{2} + \frac{h_{0}^{2}}{2} \right) \sum_{j} \sigma_{j}^{z} + J_{0}^{2} h_{z} \sum_{j} \sigma_{j}^{x} \sigma_{j+1}^{z} \sigma_{j+2}^{x} \Big] \\
- \frac{T^{3}}{24} \Big[ h_{z} J_{0} \Big(4J_{0}^{2} + 3h_{0}^{2} - 4h_{z}^{2}\Big) \sum_{j} \Big(\sigma_{j}^{x} \sigma_{j+1}^{y} + \sigma_{j}^{y} \sigma_{j+1}^{x}\Big)\\
+ h_{0} h_{z} \Big(6J_{0}^{2}+h_{0}^{2}-h_{z}^{2}\Big) \sum_{j} \sigma_{j}^{y} \\
+ 6h_{0}J_{0}^{2}h_{z} \sum_{j} \sigma_{j}^{x} \sigma_{j+1}^{y} \sigma_{j+2}^{x}\Big]
+ \mathcal{O} (T^{4}).
\end{multline}
Note that in the above Hamiltonian, the terms up to order $\mathcal{O}(T^3)$ already contain strings of spin operators longer than that present in the time-dependent charging Hamiltonian $H_c(t)$. To elaborate, terms such as $\sigma_j^x\sigma_{j+1}^y\sigma_{j+2}^x$ in $H_{c}^{F}$ (Eq.~\eqref{eq_Floquet}) couple three nearest-neighbour spins while $H_c(t)$ only has interactions between two nearest-neighbour spins. Likewise, higher order terms in Eq.~\eqref{eq_Floquet} contain even longer strings of spin operators which couple multiple spins. Thus, it is apparent that as the drive frequency is lowered, the periodic modulation tends towards a collective charging protocol. In the limit of very low frequencies, one can expect the presence of long strings of spin operators spanning the length of the battery in the  Floquet Hamiltonian, thus mimicking a global charging protocol where all the unit cells of the battery are collectively charged. In other words, the periodically driven quantum battery with the modulation defined in Eq.~\eqref{eqn_Hc_hx} can potentially lead to a quantum advantage at low drive frequencies. We also note in passing that for $h_0=0$, the Floquet Hamiltonian in Eq.~\eqref{eq_Floquet} becomes similar to the extended Ising Hamiltonian which is reducible to spinless free fermionic system having a similar decoupled structure in the momentum space as the TFIH \cite{zhang15,zhang17,bhattacharjee18}.
%However, if we set $h_{0}=0$, then the term $\sigma_j^x\sigma_{j+1}^y\sigma_{j+2}^x$ is absent in the order $\mathcal{O}(T^3)$.
%Also, the term $\sigma_j^x\sigma_{j+1}^z\sigma_{j+2}^x$ in the order $\mathcal{O}(T^2)$ does not contribute to the charging process, as it is aligned with the battery Hamiltonian (containing $\sigma_{j}^{z}$ term). Thus, unlike in the non-integrable case (i.e., $h_{0} \neq 0$), we do not find any term up to the order $\mathcal{O} (T^3)$ that can lead to collective or global charging in the integrable case (i.e., $h_{0}=0$).}

Before investigating the scaling of the charging power, it is important to ensure that the energy input from the charging Hamiltonian is extensive in the system size, which would otherwise suppress any quantum advantage \cite{farre20, rosa21}. To verify the same in our case,  we analyze the bandwidth $W$ of the charging Hamiltonian. At high frequencies, $T\to 0$, we have $H_c^F\approx h_z\sum_j\sigma_j^z$ and thus the bandwidth trivially scales linearly with system size. For small frequencies, we numerically evaluate $H_c^F$ as,
\begin{equation}\label{eq_HF}
	H_c^F = \frac{i}{T}\ln U^F,
\end{equation}
and calculate the corresponding bandwidth. As shown in Fig.~\ref{fig_bandwidth}, we find that the bandwidth at low frequency ($\omega=2$) also scales linearly with the system size. Note that the saturation observed for high $N$ is simply an artefact of the fact that the spectrum of $H_c^F$ calculated using Eq.~\eqref{eq_HF} is bounded modulo $2\pi/T$.

Finally, we plot the power, maximized over $n$, as a function of the battery size $N$ for two different frequencies in Fig.~\ref{fig_n_p}. Contrary to the expected super-linear scaling at low frequencies, we find that the power scales linearly with the system size in both the cases. In particular, we did not find the emergence of any quantum advantage for any of the frequencies within the numerically accessible range. Hence, our result demonstrates that global charging is not a sufficient condition for achieving quantum advantage, even in the case of periodically driven batteries.

\section{Summary} \label{sec_summary} 
In this work, we introduced a periodically driven quantum battery, which differs from the traditional way of charging through a static Hamiltonian. When observed at stroboscopic instants, the evolution is effectively driven by the Floquet Hamiltonian, thus leading to the possibility of engineering specific conditions that may be relatively difficult to achieve through static charging. As an immediate consequence, we have shown, by considering the periodically driven TFIH as the charging Hamiltonian that, the drive frequency can significantly alter the amount of energy stored in the battery. However, the periodically driven integrable TFIH, similar to its static counterpart, cannot provide any quantum advantage with respect to the charging power. Moving on to the case of a periodically driven non-integrable chain, we first show that the corresponding Floquet Hamiltonian consists of long string of spin operators that theoretically results in a global charging of the battery. However, no quantum advantage is manifested in the charging power in this case as well, thus corroborating previous results that global charging is only a necessary but not a sufficient condition to achieve quantum advantage.

\begin{acknowledgments}
All numerical simulations were performed using QuSpin \cite{bukov17, bukov19}. We acknowledge Amit Dutta for critical comments and suggestions. We also acknowledge Souvik Bandyopadhyay and Somnath Maity for useful discussions. SM acknowledges financial support from PMRF, MHRD, India. SB acknowledges CSIR, India for financial support.  
\end{acknowledgments}
\appendix

\section{Jordan-Wigner transformation of the transverse-field Ising Hamiltonian}\label{app_ising_integrable}
We use the following Jordan-Wigner (JW) transformations~\cite{lieb61,dutta15}, which map each spin-1/2 system to a system of spinless Fermions:
\begin{subequations}\label{eqn_JW}
	\begin{align}\label{eqn_JW1}
		c_{j}^{\dagger}=\sigma_{j}^{+} \left( \prod_{n=1}^{j-1} \sigma_{n}^{z} \right) ,
	\end{align}
	\begin{align}\label{eqn_JW2}
		c_{j}=\sigma_{j}^{-} \left( \prod_{n=1}^{j-1} \sigma_{n}^{z} \right) ,
	\end{align}
\end{subequations} 
where $\sigma_{j}^{\pm}=\frac{1}{2}(\sigma_{j}^{x} \pm i \sigma_{j}^{y})$ and $c_{j}$ ($c_{j}^{\dagger}$) is Fermionic annihilation (creation) operator for $j$-th site,  satisfying  usual anti-commutation relations. Using JW transformations as mentioned in  Eq.~\eqref{eqn_JW1} and \eqref{eqn_JW2}, the battery Hamiltonian in Eq.~\eqref{eqn_H_B} and the charging Hamiltonian $H_{c} (t) = H_{B} + V(t)$ with $V(t)$ of the form mentioned in Eq.~\eqref{eqn_Hc_Ising} are recast to the following forms:
\begin{equation}\label{eqn_JW_HB}
	H_{B} = h_{z} \sum_{j=1}^{N} ( 2 c_{j}^{\dagger} c_{j} -1 ) ,
\end{equation}
\begin{multline}\label{eqn_JW_Ising}
	H_{c} (t) = h_{z} \sum_{j=1}^{N} ( 2 c_{j}^{\dagger} c_{j} -1 ) - J (t) \sum_{j=1}^{N-1} ( c_{j}^{\dagger} c_{j+1} + c_{j+1}^{\dagger} c_{j} ) \\
	+ J(t) \sum_{j=1}^{N-1} ( c_{j} c_{j+1} + c_{j+1}^{\dagger} c_{j}^{\dagger}) .
\end{multline}

It should be noted that all  the terms in the charging Hamiltonian $H_{c}(t)$ in Eq.~\eqref{eqn_JW_Ising} are local as the Hamiltonian in Eq.~\eqref{eqn_JW_Ising} involves
only nearest neighbour interactions.

Resorting to  the Fourier space, we have,
\begin{equation}\label{eqn_HB_HBk}
	H_{B} = \bigoplus_{k \geq 0} H_{B,k} ,
\end{equation}
\begin{equation}\label{eqn_Hc_Hck}
	H_{c} (t) = \bigoplus_{k \geq 0} H_{c,k} (t) ,
\end{equation}
where $H_{B,k}$ and $H_{c,k} (t)$ with momentum
$k \in [ 0, \pi]$ are given by,
\begin{equation}\label{eqn_HB_k}
	H_{B,k} = 2 h_{z} \eta_{z} ,
\end{equation}
\begin{equation}\label{eqn_Ising_Hc_k}
	H_{c,k} (t) = 2 J(t) \sin{(k)} \eta_{y} + \left( 2 h_{z} - 2 J(t) \cos{(k)} \right) \eta_{z} ,
\end{equation}
where $\eta_{z}$ and $\eta_{y}$ are Pauli matrices representing pseudo-spins. The Floquet operator for $k$-th mode is determined by
\begin{equation}\label{eqn_Floquet_k}
	U_{k}^{F} = \mathcal{T} \exp (-i \int_{0}^{T} H_{c,k}(t) dt) = \exp (-i H_{c,k}^{F} T) ,
\end{equation}
where $H_{c,k}^{F}$ is the Floquet Hamiltonian for $k$-th mode. One can find out the  explicit form of $U_{k}^{F}$ as:
\begin{multline}\label{eqn_U_k_T_supp}
U_{k}^{F} =\exp (-i \vec{\beta}(k) \cdot \vec{\eta}) \\ =\exp \left( - \frac{iT}{2} \left( 2 J_{0} \sin{(k)} \eta_{y} + \left( 2 h_{z} - 2 J_{0} \cos{(k)} \right) \eta_{z} \right) \right) \\
\times \exp \left( - \frac{iT}{2} \left( - 2 J_{0} \sin{(k)} \eta_{y} + \left( 2 h_{z} + 2 J_{0} \cos{(k)} \right) \eta_{z} \right) \right) ,
\end{multline}
where $\vec{\beta} (k) = \beta_{x} (k) \hat{x} + \beta_{y}(k) \hat{y} + \beta_{z} (k) \hat{z}$ and $\vec{\eta}= \eta_{x} \hat{x} + \eta_{y} \hat{y} + \eta_{z} \hat{z}$.
Interestingly, using  Eq.~\eqref{eqn_U_k_T},  we note that  $U_{k = 0}^{F} = U_{k = \pi}^{F} = \mathbb{1} $, when $\omega = 2 h_{z}/p$, where $p \in \mathbb{Z}$. On the other hand,  $U_{k = 0}^{F} = U_{k = \pi}^{F} = - \mathbb{1} $ when $\omega = 4 h_{z}/(2 p +1)$, where $p \in \mathbb{Z}$. At these values of frequency, Floquet quasi-energy gap closes or reopens, respectively ~\cite{thakurathi13} (i.e., eigenvalues of the Floquet Hamiltonian become zero and $\pi$, respectively) and thus the stored energy is expected to  attain peaks~\cite{russomanno12}.

\section{Stored energy ($E$) and $\Delta H_{B}^{2}$ for periodic driving with the integrable transverse-field Ising Hamiltonian}\label{app__int}
For periodic driving with integrable transverse-field Ising Hamiltonian, the energy stored at stroboscopic time $t = n T$ is given by,
\begin{equation}\label{eqn_en}
E(n) = \sum_{k \geq 0} E_{k} (n) ,
\end{equation}
where $E_{k} (n)$ for $k \in [ 0, \pi]$ is evaluated as,
\begin{multline}\label{eqn_ekn}
E_{k}(n) = \expval{H_{B,k}}{\psi_{k} (n T)}-\expval{H_{B,k}}{\psi_{k} (0)}\\= 4 h_{z} \sin^{2} (n \beta) \left( 1- \frac{\beta_{z}^{2} (k)}{\beta^{2} (k)} \right) ,
\end{multline}
with $\vec{\beta} (k) = \beta_{x} (k) \hat{x} + \beta_{y}(k) \hat{y} + \beta_{z} (k) \hat{z}$, $\beta= |\vec{\beta} (k)|$ and the state $\ket{\psi_{k} (nT)}$ at the stroboscopic time $t = nT$ determined by,
\begin{equation}\label{eqn_psi_k_n}
\ket{\psi_{k} (nT)} = (U_{k}^{F})^{n} \ket{\psi_{k} (0)} = \exp(-i n \vec{\beta} (k)\cdot \vec{\eta}) \ket{\psi_{k} (0)} .
\end{equation}
\begin{widetext}
Similarly, the variance of the battery Hamiltonian ($\Delta H_{B}^{2}$) at time $t= nT$ is given by,
\begin{equation}\label{eqn_hbn}
\Delta H_{B}^{2} (n) = \sum_{k \geq 0} \Delta H_{B,k}^{2} (n) ,
\end{equation}
where $\Delta H_{B,k}^{2} (n)$ for $k \in [ 0, \pi]$ is determined by,
%\begin{multline}\label{eqn_hbnk}
%\Delta H_{B,k}^{2} (n) = \expval{H_{B,k}^{2}}{\psi_{k} (n T)} \\
%-(\expval{H_{B,k}}{\psi_{k} (n T)})^{2} \\= 4h_{z}^{2} - (E_{k} (n) - 2h_{z})^{2} \\=E_{k} (n) \left( 4h_{z}-E_{k} (n) \right)\\
%=16 h_{z}^{2} \sin^{2} (n \beta) \left( 1- \frac{\beta_{z}^{2} (k)}{\beta^{2} (k)} \right)\left( \cos^{2} (n \beta)+ \sin^{2} (n \beta) \frac{\beta_{z}^{2} (k)}{\beta^{2} (k)} \right) .
%\end{multline}
\begin{multline}\label{eqn_hbnk}
\Delta H_{B,k}^{2} (n) = \expval{H_{B,k}^{2}}{\psi_{k} (n T)}
-(\expval{H_{B,k}}{\psi_{k} (n T)})^{2} \\
=16 h_{z}^{2} \sin^{2} (n \beta) \left( 1- \frac{\beta_{z}^{2} (k)}{\beta^{2} (k)} \right)\left( \cos^{2} (n \beta)+ \sin^{2} (n \beta) \frac{\beta_{z}^{2} (k)}{\beta^{2} (k)} \right) =E_{k} (n) \left( 4h_{z}-E_{k} (n) \right).
\end{multline}
\end{widetext}
The variance $\Delta H_{B,k}^{2} (n)$ can also be written as,
\begin{equation}\label{eqn_hbek}
\Delta H_{B,k}^{2} (n) = 4h_{z}^{2} - (E_{k} (n) - 2h_{z})^{2}.
\end{equation}
From Eq.~\eqref{eqn_hbek}, it is clear that maximum of $E_{k} (n)$ corresponds to the minimum of $\Delta H_{B,k}^{2} (n)$, for all $k \in [0, \pi]$. This is reflected in the coincidence of the minima of the variance of the battery Hamiltonian with the maxima of the stored energy.

Further, the variance of the charging Hamiltonian ($\Delta H_{c}^{2}$) at time $t \to nT$ is given by,
\begin{equation}\label{eqn_hcn}
\Delta H_{c}^{2} (n) = \sum_{k \geq 0} \Delta H_{c,k}^{2} (n) ,
\end{equation}
\begin{widetext}
where $\Delta H_{c,k}^{2} (n)$ for $k \in [ 0, \pi]$ is given by,
\begin{multline}\label{eqn_hcnk}
\Delta H_{c,k}^{2} (n) = \expval{H_{c,k}^{2}}{\psi_{k} (n T)} -(\expval{H_{c,k}}{\psi_{k} (n T)})^{2} \\= 4J_{0}^{2} \sin^{2}(k) + \left( 2h_{z} - 2J_{0} \cos(k) \right)^{2} - \left[ \left(1- \frac{E_{k} (n)}{2} \right) \left( 2J_{0} \cos(k) - 2h_{z} \right) \right. \\ + \left. \frac{4J_{0}}{\beta (k)} \sin(k) \sin(n \beta) \left( \beta_{x} (k) \cos(n \beta)- \frac{\beta_{y} (k) \beta_{z} (k)}{\beta (k)} \sin(n \beta) \right) \right]^{2}  .
\end{multline}
\end{widetext}
%\begin{multline}\label{eqn_hcnk}
%\Delta H_{c,k}^{2} (n) = \expval{H_{c,k}^{2}}{\psi_{k} (n T)} \\
%-(\expval{H_{c,k}}{\psi_{k} (n T)})^{2} \\= 4J_{0}^{2} \sin^{2}(k) + \left( 2h_{z} - 2J_{0} \cos(k) \right)^{2}\\ - \left[ \left(1- \frac{E_{k} (n)}{2} \right) \left( 2J_{0} \cos(k) - 2h_{z} \right) \right.\\ + \left. \frac{4J_{0}}{\beta (k)} \sin(k) \sin(n \beta) \left( \beta_{x} (k) \cos(n \beta)- \frac{\beta_{y} (k) \beta_{z} (k)}{\beta (k)} \sin(n \beta) \right) \right]^{2}  .
%\end{multline}
Thus, from Eq.~\eqref{eqn_en}, Eq.~\eqref{eqn_hbn} and Eq.~\eqref{eqn_hcn}, it can be clearly seen that the stored energy, the variance of the battery Hamiltonian and the variance of the charging Hamiltonian can be written as the sum of quantities associated with the decoupled momentum modes, leading to the linear scaling with the system size $N$. Therefore, super-extensive scaling behaviours of charging power and the variance of the battery Hamiltonian are impossible for periodic driving with integrable transverse field Ising Hamiltonian.

%\section{Jordan-Wigner transformation of the transverse-field Ising Hamiltonian in presence of a longitudinal field}\label{app_ising_non-integrable}

%We use JW transformations mentioned in  Eq.~\eqref{eqn_JW1} and \eqref{eqn_JW2}, by which the charging Hamiltonian $H_{c} (t) = H_{B} + V(t)$ with $V(t)$ of the form mentioned in Eq.~\eqref{eqn_Hc_hx} is transformed into the following equation:
%\begin{multline}\label{eqn_JW_Hc_hx}
%	H_{c} (t) = h_{z} \sum_{j=1}^{N} ( 2 c_{j}^{\dagger} c_{j} -1 ) - J(t) \sum_{j=1}^{N-1} ( c_{j}^{\dagger} c_{j+1} + c_{j+1}^{\dagger} c_{j} ) \\
%	+ J(t) \sum_{j=1}^{N-1} ( c_{j} c_{j+1} + c_{j+1}^{\dagger} c_{j}^{\dagger}) \\
%	+ h_{x}(t) \sum_{j=1}^{N} \left( c_{j}^{\dagger} \prod_{n=1}^{j-1} \left( 2 c_{n}^{\dagger} c_{n} -1 \right) + \prod_{n=1}^{j-1} \left( 2 c_{n}^{\dagger} c_{n} -1 \right) c_{j} \right) ,
%\end{multline}
%In contrast to Eq.~\eqref{eqn_JW_Ising}, the term containing $h_{x} (t)$ in Eq.~\eqref{eqn_JW_Hc_hx} renders the charging Hamiltonian  non-local.

\section{Approximated Floquet Hamiltonian for periodic driving with the transverse-field Ising Hamiltonian in presence of a longitudinal field}\label{app_Floquet}
Using Magnus expansion~\cite{magnus54,mananga11}, the Floquet Hamiltonian $H_{c}^{F}$ in the high frequency limit (i.e., $\omega \gg 1$ or $T \ll 1$) is approximately found to be :
\begin{widetext}
\begin{multline}\label{eq_magnus}
H_{c}^{F} \approx \frac{1}{T} \int_{0}^{T} dt_{1} H_{c} (t_{1})
 -\frac{i}{2 T} \int_{0}^{T} dt_{1} \int_{0}^{t_{1}} dt_{2} \left[ H_{c} (t_{1}), H_{c} (t_{2}) \right]
- \frac{1}{6 T} \int_{0}^{T} dt_{1} \int_{0}^{t_{1}} dt_{2} \int_{0}^{t_{2}} dt_{3} ( \left[ H_{c} (t_{1}), \left[ H_{c} (t_{2}), H_{c} (t_{3}) \right] \right] \\
 + \left[ H_{c} (t_{3}), \left[ H_{c} (t_{2}), H_{c} (t_{1}) \right] \right] ) + \frac{i}{12 T} \int_{0}^{T} dt_{1} \int_{0}^{t_{1}} dt_{2} \int_{0}^{t_{2}} dt_{3} \int_{0}^{t_{3}} dt_{4} ([[[H_{c}(t_{1}),H_{c}(t_{2})],H_{c}(t_{3})],H_{c}(t_{4})] \\
+ [H_{c}(t_{1}),[[H_{c}(t_{2}),H_{c}(t_{3})],H_{c}(t_{4})]]
+ [H_{c}(t_{1}),[H_{c}(t_{2}),[H_{c}(t_{3}),H_{c}(t_{4})]]] 
+ [H_{c}(t_{2}),[H_{c}(t_{3}),[H_{c}(t_{4}),H_{c}(t_{1})]]]) + \mathcal{O} (T^{4}) .
\end{multline}
\end{widetext}

The transverse-field Ising Hamiltonian in presence of a longitudinal field can be recast in the following form:
\begin{equation}\label{eqn_hc_t}
H_{c} (t) = \begin{cases}
 H_{1} , & \text{for $ 0 < t < \frac{T}{2} $}\\
H_{2} , & \text{for $ \frac{T}{2} < t < T $},
\end{cases}
\end{equation}
where $$H_{1} = h_{z} \sum_{j} \sigma_{j}^{z} + J_{0} \sum_{j} \sigma_{j}^{x} \sigma_{j+1}^{x} + h_{0} \sum_{j} \sigma_{j}^{x},$$ and $$H_{2} = h_{z} \sum_{j} \sigma_{j}^{z} - J_{0} \sum_{j} \sigma_{j}^{x} \sigma_{j+1}^{x} - h_{0} \sum_{j} \sigma_{j}^{x}.$$
Using Eq.~\eqref{eqn_hc_t}, we obtain,
\begin{equation}\label{eqn_hc_t0}
\frac{1}{T} \int_{0}^{T} dt_{1} H_{c} (t_{1}) = h_{z} \sum_{j} \sigma_{j}^{z},
\end{equation}
\begin{widetext}
The term of $H_{c}^{F}$ of the order $T$ is given by the following equation:
\begin{equation}\label{eqn_hc_t1}
\frac{1}{T} \int_{0}^{T} dt_{1} \int_{0}^{t_{1}} dt_{2} \left[ H_{c} (t_{1}), H_{c} (t_{2}) \right] = i T \left( h_{z} J_{0} \sum_{j} \sigma_{j}^{x} \sigma_{j+1}^{y} + h_{z} J_{0} \sum_{j} \sigma_{j}^{y} \sigma_{j+1}^{x} + h_{z} h_{0} \sum_{j} \sigma_{j}^{y} \right).
\end{equation}
Similarly, the term of $H_{c}^{F}$ of the order $T^{2}$ is  determined by:
\begin{multline}\label{eqn_hc_t2}
 \frac{1}{T} \int_{0}^{T} dt_{1} \int_{0}^{t_{1}} dt_{2} \int_{0}^{t_{2}} dt_{3} \left( \left[ H_{c} (t_{1}), \left[ H_{c} (t_{2}), H_{c} (t_{3}) \right] \right] + \left[ H_{c} (t_{3}), \left[ H_{c} (t_{2}), H_{c} (t_{1}) \right] \right] \right) \\=  2 T^{2} \left( h_{0} J_{0} h_{z} \sum_{j} \sigma_{j}^{z} \sigma_{j+1}^{x}
+ h_{0} J_{0} h_{z} \sum_{j} \sigma_{j}^{x} \sigma_{j+1}^{z} + h_{z} ( J_{0}^{2} + \frac{h_{0}^{2}}{2} ) \sum_{j} \sigma_{j}^{z} + J_{0}^{2} h_{z} \sum_{j} \sigma_{j}^{x} \sigma_{j+1}^{z} \sigma_{j+2}^{x} \right).
\end{multline}
The term of $H_{c}^{F}$ of the order $T^{3}$ is given by the following equation:
\begin{multline}\label{eqn_hc_t3}
\frac{1}{T} \int_{0}^{T} dt_{1} \int_{0}^{t_{1}} dt_{2} \int_{0}^{t_{2}} dt_{3} \int_{0}^{t_{3}} dt_{4} ([[[H_{c}(t_{1}),H_{c}(t_{2})],H_{c}(t_{3})],H_{c}(t_{4})] \\
+ [H_{c}(t_{1}),[[H_{c}(t_{2}),H_{c}(t_{3})],H_{c}(t_{4})]]
+ [H_{c}(t_{1}),[H_{c}(t_{2}),[H_{c}(t_{3}),H_{c}(t_{4})]]] 
+ [H_{c}(t_{2}),[H_{c}(t_{3}),[H_{c}(t_{4}),H_{c}(t_{1})]]]) \\
= \frac{i T^{3}}{2} \left( h_{z} J_{0} (4J_{0}^{2} + 3h_{0}^{2} - 4h_{z}^{2}) \sum_{j} (\sigma_{j}^{x} \sigma_{j+1}^{y} + \sigma_{j}^{y} \sigma_{j+1}^{x})
+ h_{0} h_{z} (6J_{0}^{2}+h_{0}^{2}-h_{z}^{2}) \sum_{j} \sigma_{j}^{y}
+ 6h_{0}J_{0}^{2}h_{z} \sum_{j} \sigma_{j}^{x} \sigma_{j+1}^{y} \sigma_{j+2}^{x} \right) .
\end{multline}
Using Eq.~\eqref{eqn_hc_t0}, \eqref{eqn_hc_t1}, \eqref{eqn_hc_t2} and \eqref{eqn_hc_t3}, we finally obtain Eq.~\eqref{eq_Floquet} of the main text.
\end{widetext}

%\bibliographystyle{plain}
%\bibliography{mybib}

	\bibliography{reference}

%\bibitem{breuer02} H. P. Breuer,  F.  Petruccione,  Theory of open quantum systems, Oxford University Press, Oxford, (2002).
%
%\bibitem{kurizki13} D. Gelbwaser-Klimovsky, R. Alicki and G. Kurizki, Phys. Rev. E, 87, 012140 (2013).
%
%\bibitem{victor21} V. Mukherjee and U. Divakaran, J. Phys.: Condens. Matter 33, 454001 (2021).
%
%\bibitem{campaioli18} F. Campaioli, F. A. Pollock and S. Vinjanampathy, arXiv:1805.05507v1 [quant-ph] (2018).
%
%\bibitem{farre20} S. Julià-Farré, T. Salamon, A. Riera, M. N. Bera and M. Lewenstein, Phys. Rev. Research, 2, 023113 (2020).
%
%\bibitem{sourav20} S. Bhattacharjee and A. Dutta, arXiv:2008.07889v2 [quant-ph] (2020).
%
%%\bibitem{kosloff13} R. Kosloff, Entropy, 15, 2100 (2013).
%
%\bibitem{mondal20} S. Mondal, S. Bhattacharjee and A. Dutta, Phys. Rev. E, 102, 022140 (2020).
%
%\bibitem{lieb61} E. Lieb, T. Schultz, and D. Mattis, Ann. Phys. (NY) 16, 407 (1961).
%
%\bibitem{dutta15} A. Dutta, G. Aeppli, B. K. Chakrabarti, U. Divakaran, T. F. Rosenbaum, D. Sen, Quantum phase transitions in transverse field spin models (2015).
%
%\bibitem{thakurathi13} M. Thakurathi, A. A. Patel, D. Sen and A. Dutta, Phys. Rev. B, 88, 155133 (2013).
%
%\bibitem{russomanno12} A. Russomanno, A. Silva and G. E. Santoro, Phys. Rev. Lett., 109, 257201 (2012).
%
%\bibitem{weinberg17} P. Weinberg and M. Bukov, SciPost Phys. 2, 003 (2017).
%
%\bibitem{weinberg19} P. Weinberg and M. Bukov, SciPost Phys. 7, 020 (2019).

\end{document}